\documentclass[english,12pt,aps,prb,preprint]{revtex4}
\usepackage[T1]{fontenc}
\usepackage[latin9]{inputenc}
\usepackage{color}
\usepackage{graphicx}

\makeatletter
\@ifundefined{textcolor}{}
{%
 \definecolor{BLACK}{gray}{0}
 \definecolor{WHITE}{gray}{1}
 \definecolor{RED}{rgb}{1,0,0}
 \definecolor{GREEN}{rgb}{0,1,0}
 \definecolor{BLUE}{rgb}{0,0,1}
 \definecolor{CYAN}{cmyk}{1,0,0,0}
 \definecolor{MAGENTA}{cmyk}{0,1,0,0}
 \definecolor{YELLOW}{cmyk}{0,0,1,0}
 }

\makeatother

\usepackage{babel}
\begin{document}

\title{Relativistic Electron Experiment for the Undergraduate Laboratory}

\author{Robert E. Marvel}

\altaffiliation{Present address: Vanderbilt Institute of Nanoscale Science and Engineering, Vanderbilt University, Nashville, TN 37235.}

\author{Michael F. Vineyard}

\affiliation{Department of Physics and Astronomy, Union College, Schenectady,
NY 12308}
\begin{abstract}
We have developed an undergraduate laboratory experiment to make independent
measurements of the momentum and kinetic energy of relativistic electrons
from a $\beta$-source. The momentum measurements are made with a
magnetic spectrometer and a silicon surface-barrier detector is used
to measure the kinetic energy. A plot of the kinetic energy as a function
of momentum compared to the classical and relativistic predictions
clearly shows the relativistic nature of the electrons. Accurate values
for the rest mass of the electron and the speed of light are also
extracted from the data.
\end{abstract}
\maketitle

\section{Introduction}

Special relativity is a standard topic in the undergraduate physics
curriculum and one that many students find particularly interesting.
However, there are very few experiments available for the undergraduate
laboratory to illustrate relativistic effects.

Over the years several undergraduate relativity experiments have been
developed to make measurements of relativistic electrons from $\beta$-sources.\cite{bartlett,parker,geller,couch,luetzelschwab,lund}
A few others use Compton scattering of $\gamma$-rays.\cite{higbie,egelstaff,hoffman,jolivette}
Recently an experiment has been reported that uses positron annihilation.\cite{dryzek_1,dryzek_2}

In this paper we describe an undergraduate laboratory experiment in
which students make independent measurements of the momentum and kinetic
energy of relativistic electrons emitted from a $\beta$-source. A
semicircular $\beta$-ray spectrometer is used to make the momentum
measurements and the kinetic energy is measured with a solid-state
detector. The results clearly show that the relativistic relationship
must be used to describe the data. The data can also be used to determine
values for the speed of light in a vacuum and the rest mass of the
electron that agree with the accepted values within the experimental
uncertainties.

\section{Theory}

Undergraduate students learn in their introductory physics courses
that the relativistic expression for the kinetic energy $K$ of an
electron with momentum $p$ is
\begin{equation}
K=\sqrt{\left(pc\right)^{2}+\left(mc^{2}\right)^{2}}-mc^{2}\label{eq:rel_ke}
\end{equation}
\textcolor{black}{where $m$ is the rest mass of the electron and
$c$ is the speed of light in a vacuum, and that in the classical
limit this reduces to
\begin{equation}
K=\frac{p^{2}}{2m}.\label{eq:clas_ke}
\end{equation}
Clearly, independent measurements of the kinetic energy and momentum
over a range of electron momenta in the relativistic regime will distinguish
between the two expressions.}

\textcolor{black}{Rewriting the expression for the relativistic kinetic
energy in the form of the equation of a line we have
\begin{equation}
\frac{p^{2}}{2K}=\left(\frac{1}{c^{2}}\right)\left(\frac{K}{2}\right)+m.\label{eq:linear_eg}
\end{equation}
This result shows that a graph of $\frac{p^{2}}{2K}$ versus $\frac{K}{2}$
yields a line with a slope of $\frac{1}{c^{2}}$ and an intercept
of $m$. Therefore values for the speed of light and the rest mass
of the electron can be extracted by fitting a line to the data.}

In this experiment the momentum of electrons from a $\beta$-source
is measured with a semicircular $\beta$-ray spectrometer. The operation
of the spectrometer is based on the fact that \textcolor{black}{an
electron traveling with a velocity $\overrightarrow{v}$ perpendicular
to a uniform magnetic field $\overrightarrow{B}$ experiences a force
in a direction perpendicular to both the velocity and the magnetic
field and with a magnitude given by
\begin{equation}
F=evB\label{eq:force}
\end{equation}
where $e$ is the magnitude of the charge of the electron. This force
will cause the electron to travel in a circular path. Applying Newton's
second law of motion we can express the magnitude of the momentum
of the electron as
\begin{equation}
p=eBr\label{eq:momentum}
\end{equation}
 where $r$ is the radius of the circular path. Thus the momentum
can be determined from measurements of the magnetic field and the
radius of the circular path of the electrons.}

\section{Experiment}

\begin{figure}[h]
\begin{centering}
\includegraphics[width=4in]{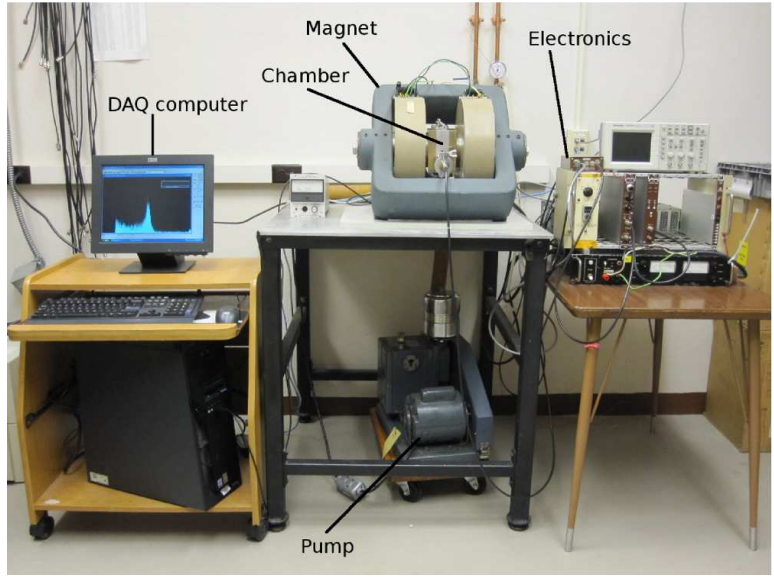}
\par\end{centering}

\caption{\label{fig:setup}(Color online) A photograph of the experimental
apparatus. The magnet, vacuum chamber, electronics, vacuum pump, and
data acquisition computer are labeled.}

\end{figure}

A photograph of the experimental setup is shown in Fig. \ref{fig:setup}.
It consists of a vacuum chamber positioned between the poles of an
electromagnet, a mechanical vacuum pump, electronics for the energy
spectrometer and magnet, and a computer for data acquisition. We will
describe the components and the experimental measurements below.

\begin{figure}[h]
\begin{centering}
\includegraphics[width=4in]{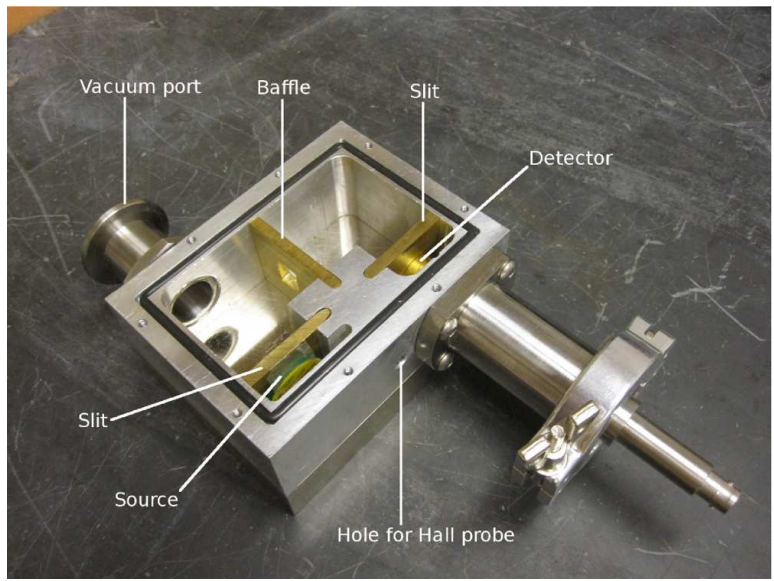}
\par\end{centering}

\caption{\label{fig:chamber_photo}(Color online) A photograph of the inside
of the vacuum chamber showing the source, slits, baffle, detector,
vacuum port, and hole for the Hall probe.}

\end{figure}

The inside of the vacuum chamber is shown in Fig. \ref{fig:chamber_photo}.
It was designed using the SolidWorks 3D CAD program\cite{solidworks}
and machined from a 12.0 $\times$ 8.9 $\times$ 3.7 cm block of 6061
aluminum. Grooves were cut on both faces for standard 1/8-inch O-rings.
The vacuum port and detector feed-through are 2.3-cm inner diameter
pipes with flanges for 1/8-inch O-rings on one end to connect to the
chamber. The connections to the vacuum hose and the BNC feed-through
for the detector are made with 1-inch Quick Flange connectors. Slits
and baffles of various sizes were cut from 3/16-inch thick brass plates.
The source is mounted behind one slit that is positioned 180$^{\circ}$
from an identical slit in front of the detector. The baffle is placed
at the 90$^{\circ}$ position. The vacuum chamber face covers are
stainless steel plates 12.0 $\times$ 8.9 $\times$ 1.9 cm with tapped
holes for bolts that go through the poles of the magnet and hold the
chamber in place. A 1/4-inch diameter hole is drilled into the outside
body of the chamber so that a Hall probe can be inserted to measure
the magnetic field between the poles of the magnet. A vacuum of around
50 mTorr is achieved using the mechanical vacuum pump.

The chamber is supported between the poles of an electromagnet with
a variable pole gap, 4-inch diameter flat pole faces, and water-cooled
coils.\cite{magnet} A 40-V, 6-A, DC power supply\cite{magnet_power_supply}
provides current for the magnet. 

\begin{figure}[h]
\begin{centering}
\includegraphics[width=4in]{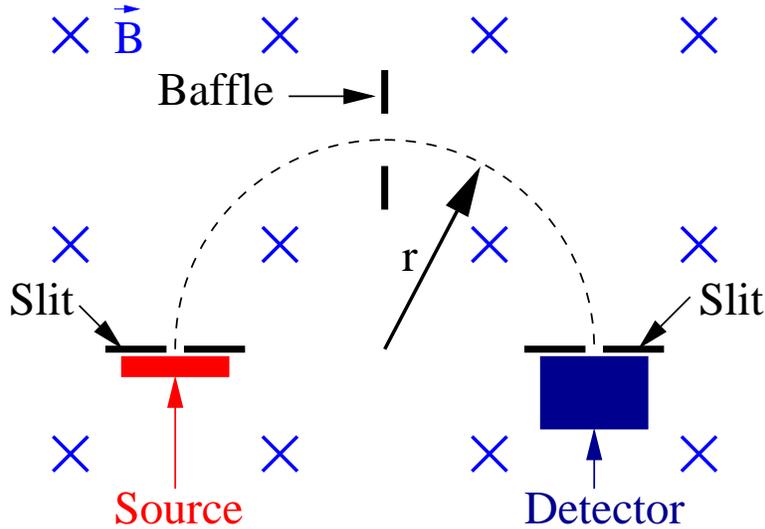}
\par\end{centering}

\caption{\label{fig:chamber_diagram} (Color online) A diagram of the inside
of the vacuum chamber showing the source, slits, baffle, detector,
magnetic field $\protect\overrightarrow{B}$, and semicircular electron
path of radius $r$.}
\end{figure}

The vacuum chamber between the poles of the magnet serves as a semicircular
$\beta$-ray spectrometer. A diagram of the spectrometer is shown
in Fig. \ref{fig:chamber_diagram}. The slits and baffle define a
semicircular path of radius $r$. The momentum of electrons from the
source that pass through the slit in front of the detector is calculated
from measurements of the radius $r$ and the magnetic field $B$.
For the data presented here we used 2 $\times$ 12 mm slits and an
11 $\times$ 12 mm baffle, and the radius $r$ = 3.24 $\pm$ 0.10
cm. This choice of slits and baffle provided a reasonable compromise
between resolution and transmission of electrons. Based on the geometry,\cite{siegbahn}
we calculated the resolution of the spectrometer to be about 3.1 \%.

\begin{figure}[h]
\begin{centering}
\includegraphics[width=4in]{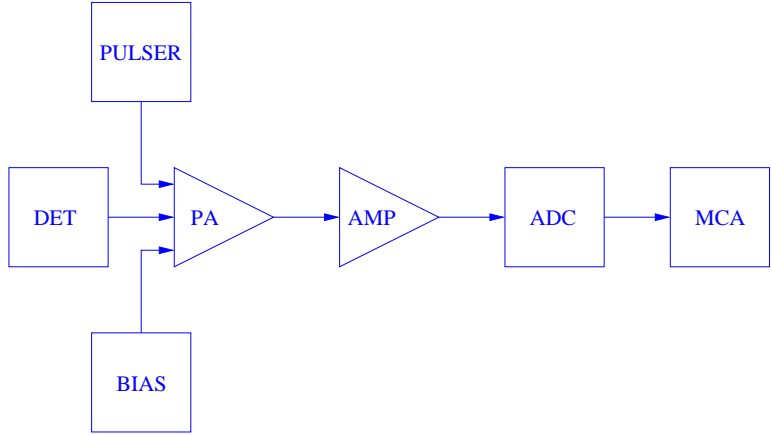}
\par\end{centering}

\caption{\label{fig:electronics}(Color online) A diagram of the electronics
used to process the signals from the silicon surface-barrier detector.}

\end{figure}

A silicon surface-barrier detector\cite{detector} with a thicknesses
of 3 mm and an active area of 25 mm$^{2}$ is used to measure the
kinetic energy of the electrons. This detector is thick enough to
stop the most energetic electrons from the source used in this experiment.
A diagram of the electronics used to process the signals from the
detector is shown in Fig. \ref{fig:electronics}. The current signal
from the detector (DET) is integrated with a preamplifier\cite{preamp}
(PA) producing a voltage pulse that is amplified with a spectroscopy
amplifier\cite{amp} (AMP) and digitized with an analog-to-digital
converter\cite{adc} (ADC). Energy spectra are created and analyzed
on the data acquisition computer using multi-channel analyzer emulator
software\cite{mca} (MCA). The bias for the detector is provided with
a 3-kV, high-voltage, power supply\cite{bias} (BIAS). We calibrate
the energy spectrometer using the 624-keV electron conversion peak
from a $^{137}$Cs source and a precision pulse generator\cite{pulser}
(PULSER).

We use a sealed, 10-$\mu$Ci, $^{204}$Tl $\beta$-source to provide
the relativistic electrons for this experiment. This source was chosen
because it has a continuous spectrum with an end-point energy of 766
keV that is well into the relativistic regime, and it is inexpensive
and readily available.\cite{source} Other suitable sources include
the continuous part of the $^{137}$Cs spectrum with an end-point
energy of 514 keV and $^{210}$Bi with an end-point energy of 1.163
MeV.

\begin{figure}[h]
\begin{centering}
\includegraphics[width=4in]{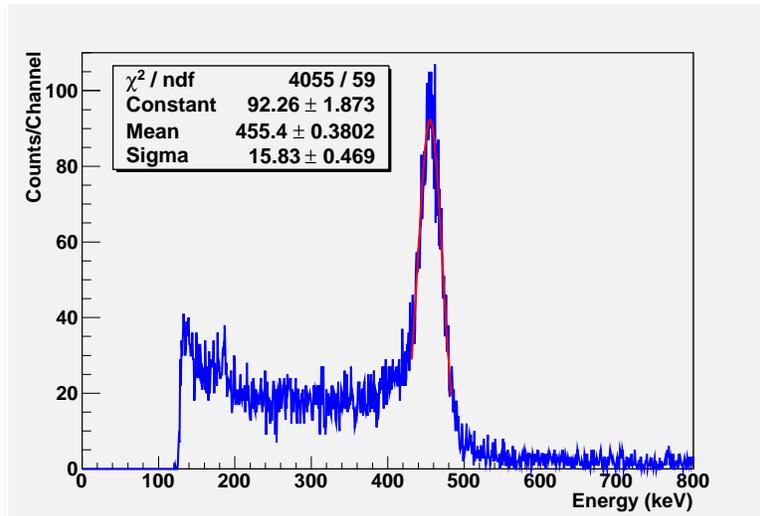}
\par\end{centering}

\caption{\label{fig:energy_spectrum}(Color online) An energy spectrum for
electrons with a momentum of 801 keV/c.}
\end{figure}

In performing the experiment, we collect electron energy spectra \textcolor{black}{for
different magnetic field settings between 500 and 1000 G (0.05 - 0.1
T). We measure the magnetic field at each setting by inserting a transverse
Hall probe connected to }a gaussmeter\cite{gaussmeter} into the small
hole drilled into the side of the vacuum chamber. The momentum of
electrons transmitted through the slit in front of the detector at
each magnetic field setting is calculated using Eq. \ref{eq:momentum}
and the measured values of the magnetic field $B$ and the radius
$r$. We fit the peak in the energy spectrum acquired at each magnetic
field setting with a Gaussian to determine the kinetic energy associated
with each momentum measurement. A typical energy spectrum for electrons
with a momentum $p$ = 801 keV is shown in Fig. \ref{fig:energy_spectrum}.
The fit to the peak yields a mean value of $K$ = 455 keV with a standard
deviation $\sigma$ = 16 keV. \textcolor{black}{Because the relative
intensity of the $\beta$ spectrum decreases with increasing energy,
the data acquisition time must be increased as the magnetic field
is increased to get sufficient statistics in each peak. }With a \textcolor{black}{10-$\mu$Ci
$^{204}$Tl source, 2-mm slits, and an 11-mm baffle, typical run times
are 2 to 12 hours per point.}

\section{Results}

\begin{figure}[h]
\begin{centering}
\includegraphics[width=4in]{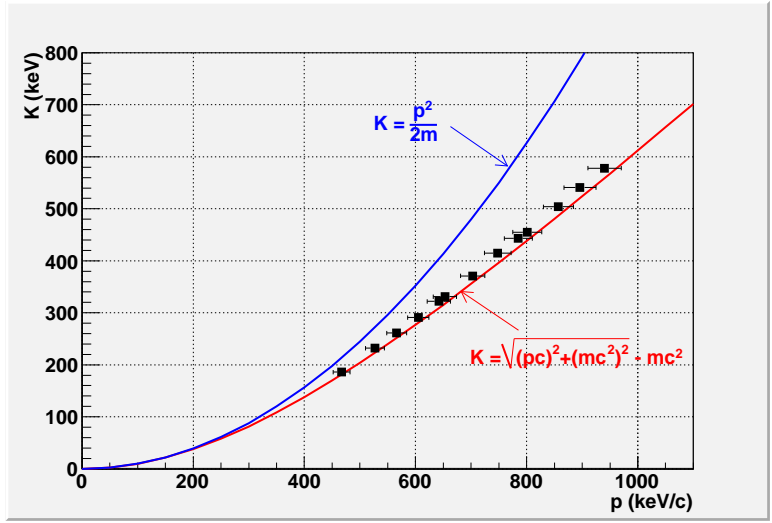}
\par\end{centering}

\caption{\label{fig:kinetic_energy_momentum}(Color online) The kinetic energy
is plotted as a function of momentum for electrons measured in this
experiment. The results of the relativistic and classical expressions
are also shown for comparison to the data.}

\end{figure}

The results for thirteen magnetic field settings between 480 and 970
G are shown in Fig. \ref{fig:kinetic_energy_momentum} where the kinetic
energy $K$ is plotted as a function of the momentum $p$. We have
assigned errors of 3.2\% on the values of momentum based on the uncertainties
in the measurements of $r$ (3.1\%) and $B$ (1\%). The uncertainties
in the measurements of the kinetic energy were at most 0.5\% resulting
in error bars that are smaller than the squares representing the data
points. The classical and relativistic relationships are shown as
curves on the figure. There is good agreement between the data and
the relativistic prediction, clearly illustrating the relativistic
nature of the electrons.

\begin{figure}[h]
\begin{centering}
\includegraphics[width=4in]{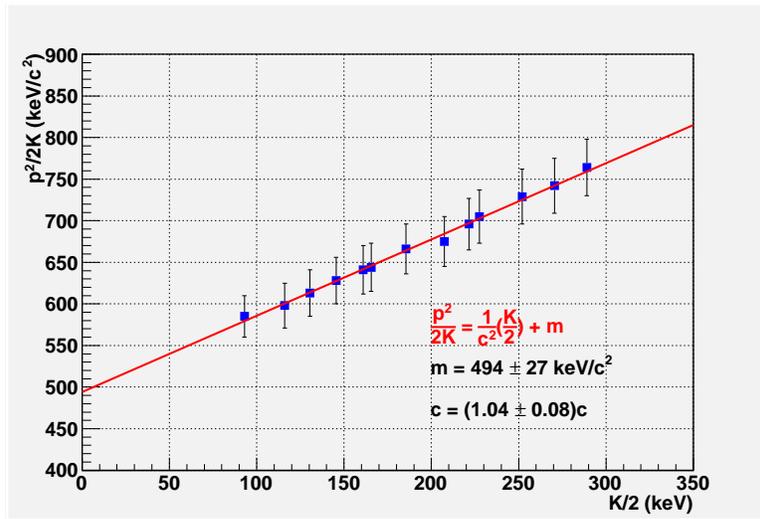}
\par\end{centering}

\caption{\label{fig:linear_plot}(Color online) A graph of the data plotted
as \textcolor{black}{$\frac{p^{2}}{2K}$ versus $\frac{K}{2}$ . The
solid line is a fit to the data used to extract values for the rest
mass of the electron and the speed of light in a vacuum.}}

\end{figure}

The data plotted as \textcolor{black}{$\frac{p^{2}}{2K}$ versus $\frac{K}{2}$
are shown in Fig. \ref{fig:linear_plot}. The line is a fit to the
data used to extract values for the rest mass of the electron and
the speed of light in a vacuum of 494 $\pm$ 27 keV/c$^{2}$ and (1.04
$\pm$ 0.08)c, respectively. These results agree with the accepted
values within the experimental uncertainties.}

\section{Summary}

We have developed an experiment in which undergraduate students make
independent measurements of the momentum and kinetic energy of electrons
from a $\beta$-source. The results clearly illustrate relativistic
effects and values for the rest mass of the electron and the speed
of light can be extracted from the data that agree with the accepted
values within the experimental uncertainties.

The first version of this experiment was developed by one of us (Marvel)
as an undergraduate senior thesis project in 2006-07 and it is now
a regular part of the advanced laboratory course at Union College.
\begin{acknowledgments}
We thank Christopher C. Jones, Emeritus Professor of Physics at Union
College, for bringing the idea for this experiment to our attention,
and John Sheehan, the technician/machinist for the Union College Department
of Physics and Astronomy, for his assistance in the design and construction
of the experimental apparatus. It is also a pleasure to thank Professor
Chad Orzel at Union College for reading a draft of this paper and
providing comments.\end{acknowledgments}

\end{document}